\documentclass[prb,floatfix,twocolumn,showpacs,aps]{revtex4}
\usepackage{graphicx}
\usepackage{dcolumn}
\usepackage{amsmath}
\usepackage{natbib}
\usepackage{amsfonts}
\usepackage{amsmath}
\usepackage{amssymb}
\usepackage{graphicx}

\input{tcilatex}

\begin{document}

\title{Anisotropic Elastic Properties of CeRhIn$_{5}$ }
\author{Ravhi S. Kumar, H. Kohlmann, B.E. Light, and A.L. Cornelius}
\affiliation{Department of Physics, University of Nevada, Las Vegas, Nevada, 89154-4002}
\author{V. Raghavan, T.W. Darling, J.L. Sarrao}
\affiliation{Materials Science and Technology Division, Los Alamos National Laboratory,
Los Alamos, NM\ 87545}

\begin{abstract}
The structure of the quasi two dimensional heavy fermion antiferromagnet
CeRhIn$_{5}$ has been investigated as a function of pressure up to 13 GPa
using a diamond anvil cell under both hydrostatic and quasihydrostatic
conditions at room ($T=295$ K) and low ($T=10$ K) temperatures.
Complementary resonant ultrasound measurements were performed to obtain the
complete elastic moduli. The bulk modulus ($B\approx 78$ GPa) and uniaxial
compressibilities ($\kappa _{a}$ $=3.96\times $ $10^{-3}$ GPa$^{-1}$ and $%
\kappa _{c}$ $=4.22\times $ $10^{-3}$ GPa$^{-1}$) found from
pressure-dependent x-ray diffraction are in good agreement with the
ultrasound measurements. Unlike doping on the Rh site where $T_{c}$
increases linearly with the ratio of the tetragonal lattice parameters $c/a,$
no such correlation is observed under pressure; instead, a double peaked
structure with a local minimum around 4-5 GPa is observed at both room and
low temperatures.
\end{abstract}

\date[Date text]{date}
\maketitle

\section{Introduction}

Ce based heavy fermion (HF) antiferromagnetic (AF) compounds have been the
subject of intensive investigations due to their unconventional magnetic and
superconducting properties. In these compounds the electronic correlations,
the magnetic ordering temperature and the crystal field effects are
sensitive to pressure, and pressure induced superconductivity has been
observed in a variety of compounds such as CePd$_{2}$Si$_{2}$, CeCu$_{2}$Ge$%
_{2}$, CeRh$_{2}$Si$_{2}$ and CeIn$_{3}$ \cite%
{Steglich79,Jaccard92,Movshovich96,Grosche96,Mathur98,Hegger00}. The
appearance of superconductivity in these systems and the deviation from
Fermi liquid behavior as a function of pressure are still challenging
problems to be studied. Recently, HF systems with the formula Ce$M$In$_{5}$ (%
$M=$ Co and Ir) have been reported to become superconductors at ambient
pressure \cite{Petrovic01,Petrovic01_2}, while CeRhIn$_{5}$ is an
antiferromagnet at ambient pressure ( $T_{N}=3.8$ K and $\gamma \thickapprox
400$ mJ/mol K$^{2}$ ). These compounds crystallizes in the HoCoGa$_{5}$
structure with alternating stacks of CeIn$_{3}$ and $M$In$_{2}$ along the $c$
axis. Thermodynamic \cite{Cornelius00}, NQR \cite{Curro00}, and neutron
scattering \cite{Bao00} experiments all show that the electronic and
magnetic properties of CeRhIn$_{5}$ are anisotropic in nature. The AF
ordering in CeRhIn$_{5}$ is suppressed with applied pressure and
superconductivity is observed at 1.6 GPa with $T_{c}=2.1$ K. Like CeIn$_{3}$%
\ the bulk nature of the SC state in CeRhIn$_{5}$ has been unambiguously
established under pressure. The AF\ state is suppressed at a pressure of
around 1.2 GPa and coexists over a limited pressure range with the
superconducting (SC) state \cite{Hegger00,Fisher02,Mito01}.

The value of $T_{c}$ in magnetically mediated superconductors is believed to
be dependent on dimensionality in addition to the characteristic spin
fluctuation temperature. Theoretical models and experimental results suggest
that SC state in CeRhIn$_{5}$ may be due to the quasi-two dimensional (2D)
structure and anisotropic AF fluctuations which are responsible for the
enhancement of\ $T_{c}$ relative to CeIn$_{3}$\cite{Pagliuso02,Monthoux01}.
A\ strong correlation between the ambient pressure $c/a$ ratio and $T_{c}$
in the Ce$M$In$_{5}$ compounds is indicative of the enhancement of the
superconducting properties by lowering dimensionality (increasing $c/a$
increases $T_{c}$) \cite{Pagliuso02}. In order to explain the evolution of
superconductivity induced by pressure and the suppression of AF ordering, it
is important to probe the effect of pressure on structure for this compound
and look for possible correlations between structural and thermodynamic
properties.

Here we report on high pressure x-ray diffraction measurements performed on
CeRhIn$_{5}$\ up to 13 GPa at high ($T=295$ K) and low ($T=10$ K)
temperatures under both hydrostatic and quasihydrostatic conditions. As the
measured linear compressibilities are similar for both the $a$ and $c$
directions, the results for all pressure measurements, both hydrostatic and
quasihydrostatic, are similar. The elastic properties obtained from the high
pressure measurements are compared to the full set of elastic constants
obtained from resonant ultrasound (RUS) measurements, and excellent
agreement is found in the measured bulk modulus ($B\approx 78$ GPa) from
both techniques. We find no direct correlation between $c/a$ and $T_{c}$ as
a function of pressure. Rather, a double peaked structure with a local
minimum around 4-5 GPa is observed for $c/a$ at both room and low
temperatures.

\section{ Experiment}

CeRhIn$_{5}$ single crystals were grown by a self flux technique \cite%
{Moshopoulou01}. The single crystals were crushed into powder and x-ray
diffraction measurements show the single phase nature of the compound. In
agreement with previous results \cite{Moshopoulou01}, the crystals were
found to have tetragonal symmetry with cell parameters $a=4.6531(1)$ \AA , $%
c=7.538(9)$ \AA .

The high pressure x-ray diffraction (XRD) experiments were performed using a
rotating anode x-ray generator (Rigaku) for Runs 1-4 ($\lambda $
=0.7093 \AA) and synchrotron x-rays
at HPCAT ($\lambda $
=0.4218 \AA), Sector 16 at the Advanced Photon Source for Run 5 and the low
temperature measurement. The sample was loaded with NaCl or MgO powder as a
pressure calibrant and either a 4:1 Methanol ethanol mixture (hydrostatic)
or NaCl (quasihydrostatic) as the pressure transmitting medium in a Re
gasket with a $180\mu $m\ diameter hole. High pressure was achieved using a
Merrill-Basset diamond anvil cell with $600$ $\mu $m culet diameters. The
XRD patterns are collected using an imaging plate ($300\times \ 300$ mm$^{2}$
) camera with $100\times \ 100$ $\mu $m$^{2}$ pixel dimensions. XRD patterns
were collected up to 13 GPa at room ($T=295$ K) and low (down to $T=10$ K)
temperatures. The low temperature measurements were made in a continuous
flow cryostat. The images were integrated using FIT2D software \cite%
{Hammersley96}. The structural refinement of the patterns was carried out
using the Rietveld method on employing the FULLPROF and REITICA (LHPM)\
software packages \cite{Rodriguez-Carvajal93}. The RUS\ technique is
described in detail elsewhere \cite{Migliori93,Migliori97}.

By measuring the resonant frequencies of a well aligned single crystal of
CeRhIn$_{5}$, we can determine the full set of room temperature elastic
constants. This will give the adiabatic bulk modulus $B^{S}$ rather than the
isothermal bulk modulus $B_{0}$ found in the pressure measurements.

\section{ Results and Discussion}

\begin{figure}[tbp]
\includegraphics[width=3.5in]{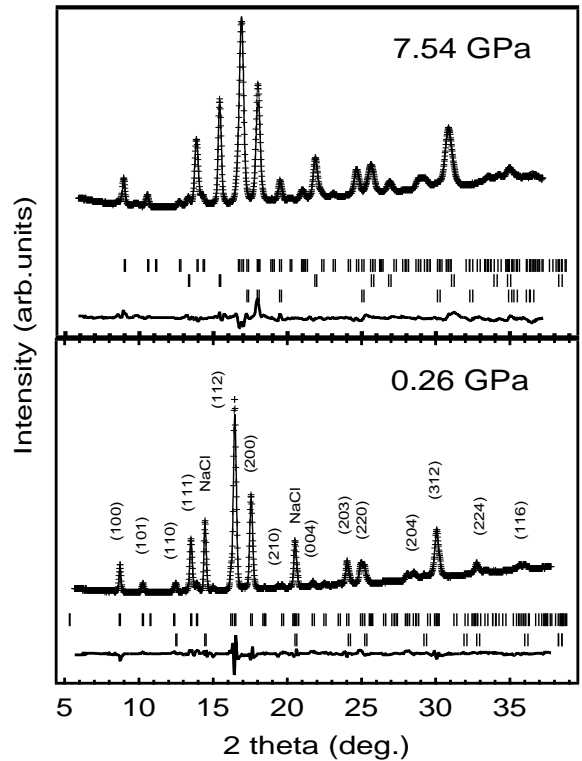}
\caption{Rietveld refinements for the high pressure x-ray diffraction
patterns of CeRhIn$_{5}$ at 0.26 GPa and 7.54 GPa. The NaCl pressure marker
and various reflections from CeRhIn$_{5}$ are labeled.}
\label{xrd}
\end{figure}
\begin{figure}[tbp]
\includegraphics[width=3.5in]{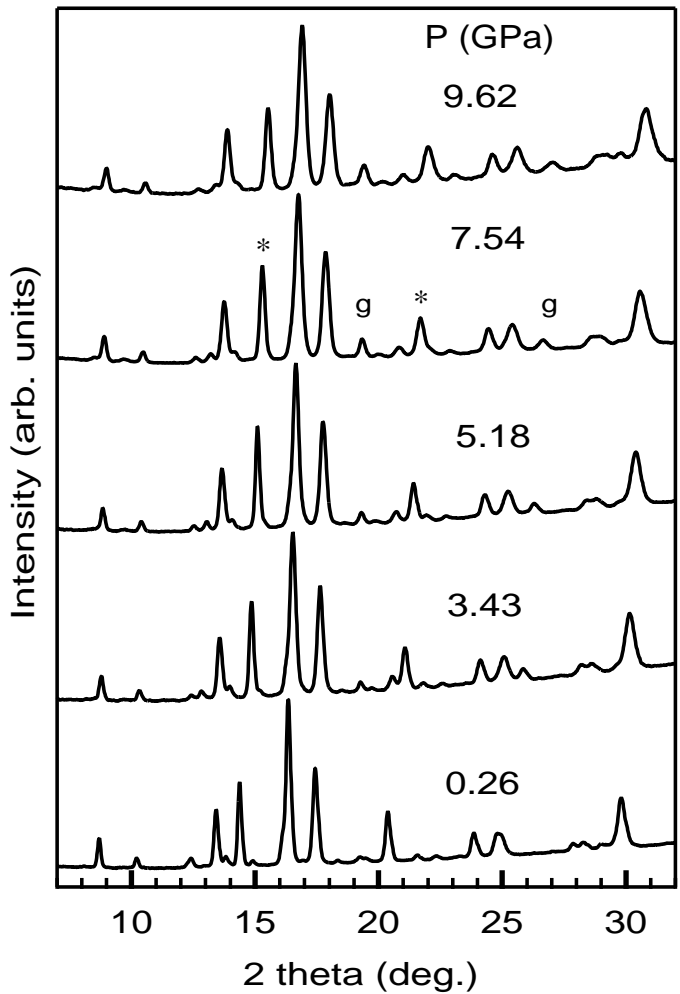}
\caption{X-ray diffraction patterns collected at various pressures for CeRhIn%
$_{5}$. Peaks labeled (g) correspond to the Re gasket and (*)\ to the NaCl
pressure marker. Peaks without a label are from the CeRhIn$_{5}$ sample.}
\label{xrd2}
\end{figure}

In Fig. \ref{xrd} we show the XRD patterns for CeRhIn$_{5}$ obtained at two
different quasihydrostatic pressures with NaCl used as the pressure
transmitting media. The raw data (crosses), Rietveld fit to the data (solid
line through the data points), fit reflections (vertical lines) and the
difference between the fit and experiment (solid line near bottom) are all
shown. Fig. \ref{xrd2} 
shows the diffraction data at five different pressures. Diffraction peaks
from the Re gasket (labeled g), NaCl (labeled *) and CeRhIn$_{5}$ (no label)
are all observed in Fig. \ref{xrd2}. The known equation of state for NaCl 
\cite{Brown99} or the standard ruby fluorescence technique \cite%
{Piermarini75} was used to determine the pressure. The refinement of the
CeRhIn$_{5}$ XRD patterns was performed on the basis of the P4/mmm space
group (No. 123). The HoCoGa$_{5}$ structure in which CeRhIn$_{5}$
crystallizes contains layers of cubo-octohedra of the structural type of AuCu%
$_{3}$ and layers of PtHg$_{2}$ structure type. The unit cell consists of Ce
atoms situated at the corners and In atoms at two inequivalent sites. In1 is
surrounded by Ce and located at the top and bottom faces while In2 is
stacked between Ce-In and Rh layers. The hybrid structure is related to both
CeIn$_{3}$ and Ce$_{2}$RhIn$_{8}$. When comparing the crystallographic data
and bulk modulus of CeIn$_{3}$ it is evident that the Ce atom in CeRhIn$_{5}$
experiences a chemical pressure of 1.4 GPa at ambient conditions \cite%
{Hegger00,Cornelius00}.

The results of the Rietveld refinement \ at different pressures have been
listed in Table \ref{Refinement}.

\begin{table}[tbp]
\centering  
\begin{tabular}{lllll}
& 1.47 GPa & 3.97 GPa & 5.18 GPa & 7.54 GPa \\ 
$a$(\AA ) & 4.6263(3) & 4.5718(3) & 4.5712(3) & 4.5298(3) \\ 
$c$(\AA ) & 7.505(1) & 7.409(1) & 7.396(1) & 7.337(1) \\ 
In2 ($z$) & 0.3036(3) & 0.3049(4) & 0.3089(3) & 0.3058(3) \\ 
$B_{\text{Ce}}$(\AA $^{2}$) & 0.3(1) & 0.5(2) & 0.3(1) & 0.5(1) \\ 
$B_{\text{Rh}}$(\AA $^{2}$) & 0.9(1) & 1.3(3) & 1.0(2) & 1.7(1) \\ 
$B_{\text{In1}}$(\AA $^{2}$) & 1.7(2) & 3.8(4) & 1.4(3) & 1.4(2) \\ 
$B_{\text{In2}}$(\AA $^{2}$) & 1.4(1) & 1.3(1) & 1.2(1) & 0.80(7) \\ 
$R_{p}$ (\%) & 2.1 & 2.6 & 2.5 & 2.1 \\ 
$R_{wp}$ (\%) & 3.0 & 3.8 & 3.5 & 3.0%
\end{tabular}%
\caption{Room temperature structural parameters, isotropic thermal
parameters $B$, and $R$ factors of CeRhIn$_{5}$ at different pressures. The
crystal structure is tetragonal and space group symmetry is P4/mmm (No.123)
with Z = 1. The atomic sites are Ce at 1a [ 0, 0, 0 ], Rh at 1b [ 0, 0,0.5
], In1 at 1c [ 0.5, 0.5, 0] and In2 at 4i [0, 0.5, $z$].}
\label{Refinement}
\end{table}
\begin{figure}[tbp]
\includegraphics[width=3.2in] {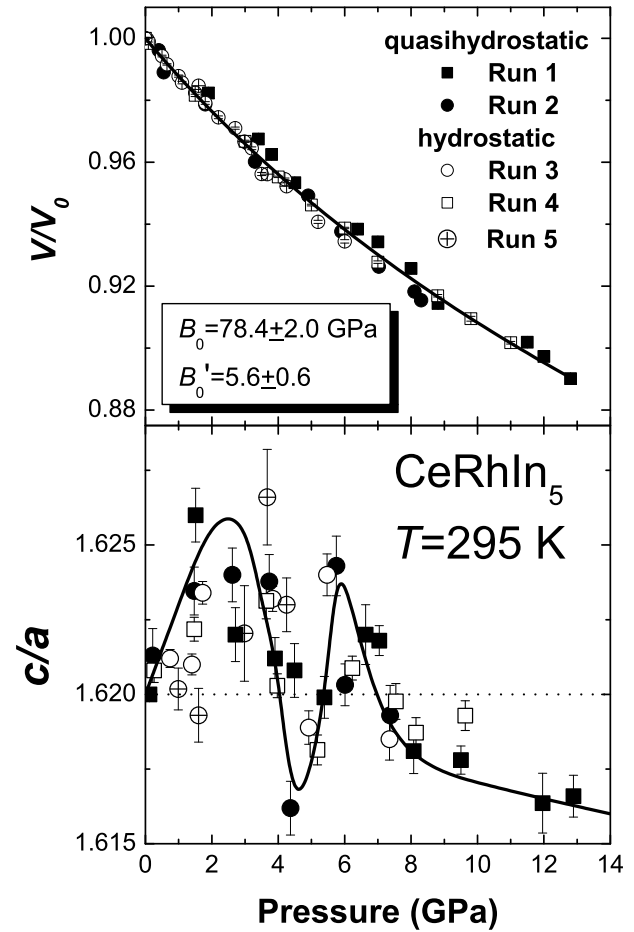}%
\caption{Normalized volume $V/V_{0}$ and ratio of tetraganol lattice
constants $c/a$ plotted versus pressure for CeRhIn$_{5}$ at room
temperature. Five separate runs, two quasihydrostatic (filled symbols) and
three hydrostatic (open symbols) are displayed. The solid line through the
volume data is a fit to all of the data using $B_{0}=78.4$ GPa and $%
B_{0}^{\prime }=5.6$. The dashed line shows the ambient pressure $c/a$
value. The line through the $c/a$ data is a guide for the eye.}
\label{compression}
\end{figure}

During the refinement, a total of nineteen parameters have been optimized
which include the background, scale factors, lattice parameters, profile
parameters, temperature factors, zero point shift parameter and atomic
coordinate. Initially the refinement has been started for two phases in most
cases including the pressure calibrant, and at higher pressures an
additional phase for the gasket has been added. At higher pressures,
considerable changes in the isotropic temperature factors are observed for
In1, In2 and \ Rh during the refinement.

The $V(P)$ data has been plotted for CeRhIn$_{5}$ for quasihydrostatic (Run
1 and Run 2) and hydrostatic (Runs 3-5) measurements in Fig. \ref%
{compression}. Since the maximum volume compression is only of the order of
10\%, the $V(P)$ data has been fit using a least squares fitting procedure
to the second order Murnaghan equation of state%
\begin{equation}
P=\frac{B_{0}}{B_{0}^{\prime }}\left[ \left( \frac{V_{0}}{V(P)}\right)
^{B_{0}^{\prime }}-1\right] .
\end{equation}%
For the room temperature ($T=295$ K) data in Fig. \ref{compression}, we find 
$B_{0}=78.4\pm 2.0$ GPa and $B_{0}^{\prime }=5.6\pm 0.6$. The RhIn$_{2}$
layers in CeRhIn$_{5}$ appear to stiffen the structure relative to CeIn$_{3}$
which has a smaller bulk modulus ($B=67$ GPa) \cite{Vedel87}. The bulk
modulus value compares well with the values reported for other HF systems 
\cite{Penney82,Spain86,Kutty87,Wassilew-Reul97}. Fig. \ref{compression} also
shows the ratio of the lattice constants $c/a$ as a function of pressure.
For all of the measurements, there appears to be a double peak structure
with a local minimum around 4-5 GPa. Note that the istropic thermal
paramaters for the In sites, in particular the In1 site, have their largest
values around 4 GPa. The initial values of the linear compressibilities
(average values from the hydrostatic measurements for $P<2$ GPa) are $\kappa
_{a}$ $=(3.96\pm 0.08)\times $ $10^{-3}$ GPa$^{-1}$ and $\kappa _{c}$ $%
=(4.22\pm 0.10)\times $ $10^{-3}$ GPa$^{-1}$. The similarity between the
measured values of $\kappa _{a}$ and $\kappa _{b}$ are likely the reason
that no discernible difference is found for the hydrostatic and
quasihydrostatic cases. The $P-V$ data shows that the system retains its
crystal structure up to the pressure limit (13 GPa) investigated.

\begin{figure}[tbp]
\includegraphics[width=3.2in] {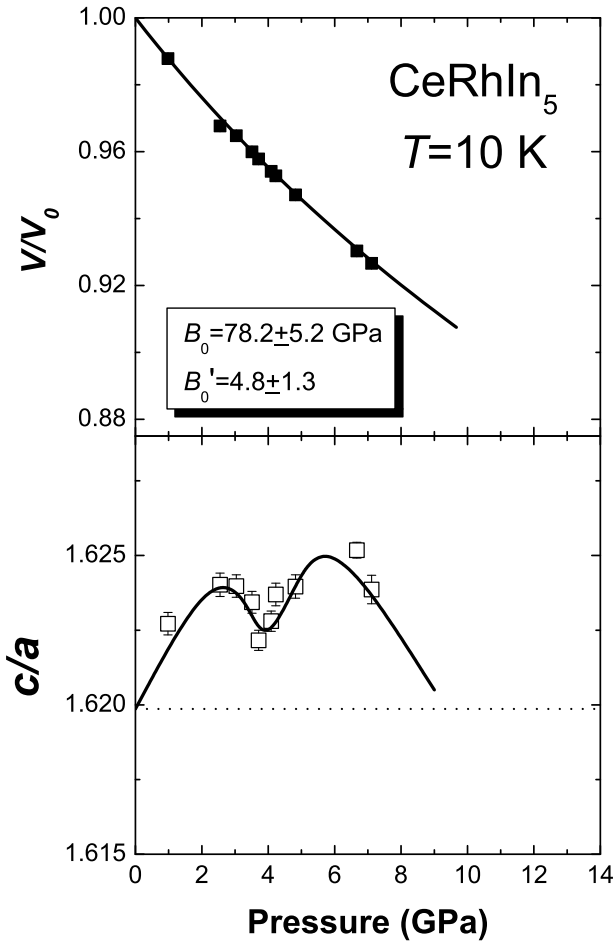}%
\caption{Normalized volume $V/V_{0}$ and ratio of tetraganol lattice
constants $c/a$ plotted versus pressure for CeRhIn$_{5}$ at 10 K. The solid
line through the volume data is a fit to all of the data using $B_{0}=78.2$
GPa and $B_{0}^{\prime }=4.8$. The dashed line shows the ambient pressure $%
c/a$ value. The solid line through the $c/a$ data is a guide for the eye. }
\label{plowt}
\end{figure}

We have also investigated the $V(P)$ behavior at low temperature $(T=10$ K).
As the superconducting transition has a maximum around 2 K, it is desirable
to obtain structural data in the low temperature regime when trying to
correlate superconductivity to structural measurements. The results for a
single hydrostatic measurement at 10 K\ is shown in Fig. \ref{plowt}. 
The value of $B_{0}=78.2\pm 5.2$ GPa is identical to the room temperature
value within the experimental uncertainty. Though the lattice does contract
at ambient pressure as temperature is lowered which would lead to a higher
bulk modulus, the expected change is within our experimental uncertainty.
The variation of $c/a$ as a function of pressure again shows a double
maximum structure at low temperature in a manner similar to the room
temperature data.

\begin{figure}[tbp]
\includegraphics[width=3.2in] {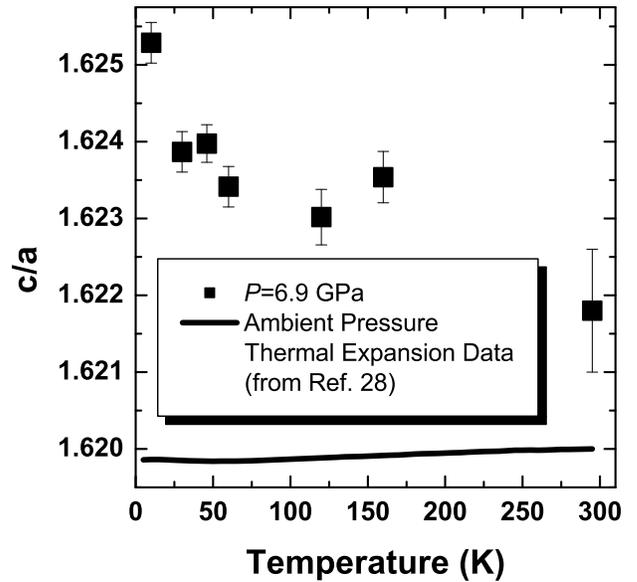}%
\caption{Ratio of tetragonal lattice constants $c/a$ plotted versus
temperature for CeRhIn$_{5}$ at $P=6.9$ GPa. The solid line is from ambient
pressure thermal expansion data (Ref. 28).}
\label{coavst}
\end{figure}

As mentioned, a\ strong correlation between the ambient pressure $c/a$ ratio
and $T_{c}$ in the Ce$M$In$_{5}$ compounds has been observed (increasing $c/a
$ increases $T_{c}$) \cite{Pagliuso02}. To further investigate the variation
of $c/a$ with pressure and temperature, we plot the value of $c/a$ as a
function of temperature at $P=6.9$ GPa in Fig. \ref{coavst}. 
As can be seen, there is a significant enhancement of $c/a$ at 6.9 GPa
relative to the ambient pressure thermal expansion data of Takeuchi \textit{%
et al}.\cite{Takeuchi01} Unlike the ambient pressure data, $c/a$ appears to
increase as temperature is lowered at 6.9 GPa. Taken as a whole, the current
results seem to show no direct correlation between the values of $c/a$ and $%
T_{c}$ under pressure. However, the value of $c/a$ ($\approx 1.624$) where $%
T_{c}(P)$ has its maximum around 2.5 GPa is consistent with a correlation
between the room temperature value of $c/a$ and $T_{c}$ for various Ce$M$In$%
_{5}$ compounds.\cite{Pagliuso02} This leads to the natural conclusion that
hybridization effects are likely the driving force behind the observed $%
T_{c}\left( P\right) $ behavior in CeRhIn$_{5}$. We will discuss this in
further detail later.

The complete set of elastic constants were measured using the RUS\ technique
and the results are shown in Table \ref{rusdata}. 
\begin{table}[tbp]
\centering                                
\begin{tabular}{|c|c|}
\hline
\textbf{Elastic Constant} & \textbf{Value (GPa)} \\ \hline
$C_{11}$ & 146.7 \\ \hline
$C_{12}$ & 45.8 \\ \hline
$C_{44}$ & 43.4 \\ \hline
$C_{33}$ & 141.4 \\ \hline
$C_{13}$ & 54.0 \\ \hline
$C_{66}$ & 41.8 \\ \hline
\textbf{Moduli} & \textbf{Value (GPa)} \\ \hline
$B^{S}$ (RUS) & 82.5 \\ \hline
$C^{t}$ (RUS) & 43.2 \\ \hline
$B_{0}$ ($P$) & 78.4 \\ \hline
\textbf{Compressibilities} & \textbf{Value (GPa$^{-1}$)} \\ \hline
$\kappa _{a}$ (RUS) & $4.09\times $ $10^{-3}$ \\ \hline
$\kappa _{c}$ (RUS) & $3.96\times $ $10^{-3}$ \\ \hline
$\kappa _{a}$ ($P$) & $3.96\times $ $10^{-3}$ \\ \hline
$\kappa _{c}$ ($P$) & $4.22\times $ $10^{-3}$ \\ \hline
\end{tabular}%
\caption{A summary of the CeRhIn$_{5}$ elastic constants $C_{ij}$ measured
using resonant ultrasound and the various moduli and compressibilities
measured by resonant ultrasound (RUS) and pressure ($P$).}
\label{rusdata}
\end{table}
The values of the adiabatic compressibility $B^{S},$ tetragonal shear
modulus $C^{t}$, and linear compressibilities ($\kappa _{a}$, $\kappa _{c}$)
can be calculated from the measured elastic constants \cite{Boettger97} and
are given by 
\begin{equation}
B^{S}=\frac{C_{33}(C_{11}+C_{12})-2C_{13}^{2}}{2C_{33}+C_{11}+C_{12}-4C_{13}}
\end{equation}%
and

\begin{equation}
C^{t}=\frac{1}{6}\left( 2C_{33}+C_{11}+C_{12}-4C_{13}\right) .
\end{equation}%
The results are displayed in Table \ref{rusdata}. The value of $B^{S}$ is
slightly larger than the isothermal value $B_{0}$ obtained from the pressure
measurements. This is to be expected as the ratio $B^{S}/B_{0}=1+\beta
\gamma _{th}T$, where $\beta =4.6\times $ $10^{-5}$ K$^{-1}$ is the volume
thermal expansion coefficient \cite{Takeuchi01} and $\gamma _{th}$ is the
thermal Gruneisen parameter which is typically of the order of unity. At
room temperature then, one then expects $B^{S}/B_{0}\approx 1.01-1.02$ which
is in reasonable agreement with our experimental value of $1.05\pm 0.03$.

In all of the measurements, the $c/a$ ratio is found to have a double peaked
structure. As mentioned previously, the hybridization between the Ce 4f
electrons and the conduction electrons should mainly depend on the distance
between Ce and its nearest neighbors. In fact, a simple model to estimate
the hybridization by means of a tight-binding calculation shows that the
hybridization should have the relatively strong $d^{-6}$ dependence for
hybridization between f and d electrons, where $d$ is the distance between
the atoms containing the d and f electrons (in our case, this would be Rh
and Ce respectively) \cite{Harrison83,Endstra93}.

\begin{figure}[tbp]
\includegraphics[width=3.5in]{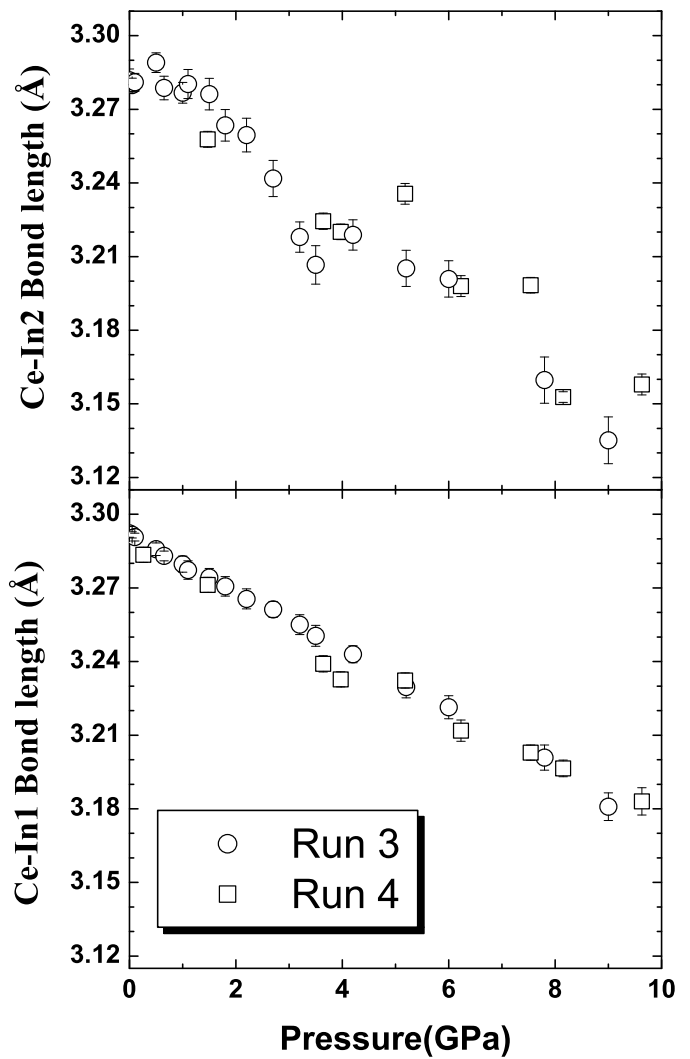}%
\caption{The measured Ce-In1 and Ce-In2 bond lengths for both hydrostatic
measurements on CeRhIn$_{5}$ as a function of pressure.}
\label{cein2}
\end{figure}

To examine the pressure dependence of $d$, the Ce-In1 and Ce-In2 bond
lengths are plotted in Fig. \ref{cein2} for the hydrostatic measurements.
The Ce-In1 bond is less compressible than the Ce-In2 bond. The Ce-In2\ data
appears to display plateaus between 0-2 and 3-5 GPa. The structural results
may be compared with the high pressure resistivity experimental data
reported for CeRhIn$_{5}$ \cite{Hegger00,Muramatsu01}. First, the
temperature corresponding to the maximum in the resistivity, often taken to
be a measure of the Kondo temperature $T_{K}$ is seen to initially decrease
in CeRhIn$_{5}$ in contrast to the usually observed behavior \cite%
{Thompson93}. One possible explanation for this effect could lie in an
initial increase in the Ce-In2 bond length causing an anomalous initial
decrease in the hybridization. Whereas the plot of Ce-In1 bond length with
pressure shows a gradual decrease with increasing pressure. Our data is not
sufficient to make any definite conclusions along these lines. The smooth
decrease in the Ce-In1\ bond length would lead one to expect the typical
inverse parabolic $T_{c}(P)$ dependence consistent with theoretical
calculations \cite{Monthoux01}, measurements on CeRhIn$_{5}$ \cite%
{Hegger00,Muramatsu01} and most heavy fermion superconductors \cite%
{Movshovich96,Grosche96,Mathur98}.

\section{Conclusions}

We have studied the elastic properties of the heavy fermion system CeRhIn$%
_{5}$ using resonant ultrasound and hydrostatic and quasihydrostatic
pressures up to 13 GPa using x-ray diffraction. The bulk modulus ($B=78$
GPa) and uniaxial compressibilities ($\kappa _{a}$ $=3.96\times $ $10^{-3}$
GPa$^{-1}$ and $\kappa _{c}$ $=4.22\times $ $10^{-3}$ GPa$^{-1}$) found from
pressure-dependent x-ray diffraction are in good agreement with the
ultrasound measurements. Unlike doping experiments which hint at a strong
correlation between the $c/a$ ratio and $T_{c}$, pressure shows no such
correlation as a double peaked structure with a local minimum around 4-5 GPa
is found at 295 K and 10 K.

\begin{acknowledgments}
We thank Maddury Somayazulu, Beam Line Scientist at HPCAT for assistance on the low temperature, high pressure diffraction measurements. Work at UNLV is supported by DOE EPSCoR-State/National Laboratory
Partnership Award DE-FG02-00ER45835. Work at LANL is performed under the
auspices of the U.S. Department of Energy. HPCAT is a collaboration among
the UNLV High Pressure Science and Engineering Center, the Lawrence
Livermore National Laboratory, the Geophysical Laboratory of the Carnegie
Institution of Washington, and the University of Hawaii at Manoa. The UNLV
High Pressure Science and Engineering Center was supported by the U.S.
Department of Energy, National Nuclear Security Administration, under
Cooperative Agreement DE-FC08-01NV14049. Use of the Advanced Photon Source
was supported by the U. S. Department of Energy, Office of Science, Office
of Basic Energy Sciences, under Contract No. W-31-109-Eng-38.
\end{acknowledgments}

\newif\ifabfull\abfulltrue

\end{document}